Muhammad FAROOQ-I-AZAM[1], Muhammad Naeem AYYAZ[2], Saleem AKHTAR[3]

COMSATS Institute of Information Technology Lahore (1) (3), University of Engineering & Technology Lahore (2)

# Connectivity based technique for localization of nodes in wireless sensor networks

*Abstract. We propose a localization algorithm for wireless sensor networks, which is simple in design, does not involve significant overhead and yet provides acceptable position estimates of sensor nodes. The algorithm uses settled nodes as beacon nodes so as to increase the number of beacon nodes. The algorithm is range free and does not need any additional piece of hardware for ranging. It also does not involve any significant communication overhead for localization. The simulation and results show that good localization accuracy is achieved for outdoor environments.*

*Streszczenie. In this place the editor of journal inserts Polish version of the abstract. Please leave three lines for this abstract. Of course Polish language Authors are requested to prepare also Polish "Streszczenie". All papers should have two sets: title, abstract, keywords - Polish and English. (Przygotowanie artykułu do Przeglądu Elektrotechnicznego - polski tytuł na końcu streszczenia - Polish tittle at the end).*
x

**Keywords:** localization; positioning; location estimation; position estimation; wireless sensor networks.

**Introduction**

Certain applications of wireless sensor networks require that the sensor nodes should be aware of their positions in the sensor field. To be significant and meaningful, the data such as temperature, humidity and pressure gathered by sensor nodes must be ascribed to the position from where it was collected. For this to happen, the sensor nodes must find their positions. The literature has come to term this problem of location or position estimation of sensor nodes simply as *localization*. Under certain circumstances, the nodes should not only by aware of their position but also the direction or orientation relative to the network [1].

Let us now consider a sensor network which is symmetric, two-dimensional and arranged in a square shape. Then this sensor network can be represented as a graph $G(V, E)$ where the set of sensor nodes can be represented as set of vertices as under:

(1) $\quad V = \{v_1, v_2, ..., v_n\}$

The set of edges $E$ in the graph $G(V, E)$ comprises of all edges $e = (i, j) \in E$ iff $v_i$ reaches $v_j$ i.e. the distance between $v_i$ and $v_j$ is less than $r$ where $r$ is the maximum distance between the two nodes after which communication between them ceases to exist i.e. if the distance between two nodes is greater than $r$, no direct communication between them is possible. Put in other words, if the distance between any two nodes is greater than $r$, the two nodes are not *neighbor* nodes. The distance between two neighbor nodes $v_i$ and $v_j$ is defined as the weight $w(e) \leq r$ of the edge $e = (i, j)$ between them.

The sensor node localization problem can now be stated as following:

*Let there be a multihop sensor network represented by a graph $G = (V, E)$. The graph has a set of beacon nodes $B$ with known positions given by $(x_b, y_b)$ for all $b \in B$. The localization problem requires to find the position set $(x_d, y_d)$ of as many dumb nodes $d \in D$ as possible. Finding the location of a node implies finding its latitude, longitude and altitude.*

**Related work**

A number of localization schemes for single-hop networks have been developed. Some of the localization schemes for single-hop networks include active badge [2], active office [3, 4], cricket [5, 6, 7], Bulusu's algorithm [8] and APIT [9]. Majority of these single-hop localization algorithms were developed for context-aware computing before research on wireless sensor networks gained focus.

One of the localization schemes which determines positions of dumb nodes using only connectivity information is due to Bulusu, Heidmann and Estrin [8]. Each node transmits periodic beacon signals every $T$ seconds containing its position information. Each dumb node $j$ keeps a count of the number of beacon signals received from a particular beacon node $i$ in some fixed time interval $t$. Knowing the time period $T$ after which a beacon signal is transmitted by the beacon node $i$, the dumb node $j$ can also compute the total number of beacon signals transmitted by the beacon node $i$ in time interval $t$. Using both these parameters, the dumb node $j$ can compute a connectivity metric for a particular beacon node. The connectivity metric $CM_{ij}$ is given by the percentage of beacon signals received by the dumb node $j$ which were transmitted by the beacon node $i$. Higher is the value of $CM_{ij}$, greater is the number of $B_i$ beacon signals received by the dumb node $D_j$, and smaller is the distance between them. The dumb node $D_j$ calculates this connectivity metric for all the beacon nodes in the set $S$ comprising of all beacons nodes from which it receives beacons. From this set $S$ of beacon nodes, it selects a subset $N$ of those neighbor beacon nodes for which the connectivity metric exceeds a certain threshold, say, 90 percent. The dumb node $D_j$ then localizes itself by determining the centroid of the selected beacon nodes. Our localization scheme presented in the next section also uses connectivity information for localization of dumb nodes and derives from Bulusu's work.

Another localization scheme is APIT, which is an area-based range-free localization scheme proposed by He et al. [9]. Rather than context-aware applications, APIT algorithm is designed for sensor networks. APIT uses beacon nodes and RSSI information from neighbor nodes of a dumb node to solve the problem of localization. It employs distributed processing and each dumb node determines its position by locally processing the available information.

In addition to the algorithms described above, a number of localization algorithms have been proposed such as Recursive Position Estimation [10] and Directed Position Estimation [1]. Recursive Position Estimation (RPE), as the name suggests, utilizes recursive steps to estimate and refine the position estimation. Although its communication cost is low compared to other localization algorithms, it needs relatively more beacon nodes. Directed Position

Estimation (DPE) algorithm is like RPE but reduces the required number of beacon nodes by using directed recursion. Another localization scheme proposed by De Oliveira et al. [11] combines the problem of localization with time synchronization in sensor networks and gives an algorithm which solves both the problems simultaneously. Doherty, Pister & Ghaoui [12] treat the problem of localization as an optimization problem and applies linear programming to arrive at the solution of location problem. Another approach [13] uses mobile beacon nodes for localization of dumb nodes and employs distance estimation with weighted least squares. Some other localization schemes include Active Badge [2], Active Office [3, 4], Ad hoc Positioning System [14], Cricket [5, 6, 7], iterative localization [15, 16], collaborative multilateration [17, 18] and Multidimensional Scaling [19, 20, 21]. Active Office, Active Badge and Cricket are meant for context-aware applications and are not suitable for wireless sensor networks. Ad hoc Positioning System and Multidimensional Scaling provide good academic insight into the problem of localization but are not practicable due to their complex nature. Iterative and collaborative multilateration are only frameworks that focus on beacon node redundancy and localization of blind nodes. However, the principles outlined in these frameworks can be utilized to build better localization algorithms.

Examples of some recent localization algorithms are the ones proposed by Chang, Hung, Lin, & Li [22] and Wang, Wu & Shu [23]. The algorithm proposed in [22] uses mobile nodes to ascertain the location of dumb nodes and it only provides coarse-grained location estimates. The authors in [23] describe a particle filter to improve the reliability of RSS measurements. This implies cost overhead in terms of energy and processing which are usually not easily available in sensor networks.

**Proposed localization algorithm**

For the purpose of description of the algorithm, let's assume that there is a sensor field deployed with beacon and dumb nodes. All beacon and dumb nodes have identical and spherical radio range in a radius $r$. Sensor network is deployed in an outdoor unconstrained environment such that all nodes have line of sight communication.

Each beacon node transmits beacon signals every $T$ seconds where $T$ is fixed and is same for all beacon nodes. In other words, the time interval between two successive transmissions of a beacon signal by any beacon node is $T$. Neighbor beacon nodes transmit beacon signals in such a way that their beacon signals are not transmitted concurrently. This can be achieved during the bootstrap when the beacon nodes are deployed. During this period, the beacon nodes decide upon the sequence of beacon transmissions of neighbor beacon nodes. It is assumed that some collision avoidance mechanism is already being used by the sensor node at data link layer.

After the beacon nodes have been deployed and they start transmitting beacons every $T$ seconds, each dumb node maintains a table of all beacon nodes from which it receive beacon signals. A dumb node $D$ also keeps a count of the number of beacon signals it receives from a particular beacon node $B$. Furthermore, the dumb node $D$ counts the number $N_b$ of beacon signals received from a particular beacon node for a sampling time period $T_s$. With the knowledge of time period of transmission of beacon signals, the dumb node also computes the number $N_B$ of total beacon messages transmitted by the beacon node in the sampling time period $T_s$.

After computing the values of $N_b$ and $N_B$, the dumb node calculates proximity factor $F_B$ of the particular beacon node:

$$(2) \quad F_B = \frac{100 \times N_b}{N_B}$$

The dumb node calculates the proximity factor for each beacon node from which it receives beacon signals and stores it in the table of beacon nodes it maintains. This step is repeated by each dumb node for all the beacon nodes from which it receives communication.

Next, the dumb node prepares a list of threshold values of the proximity factor. The dumb node $D$ then selects 3 or more beacon nodes whose proximity factor falls in the highest range i.e. greater than the highest threshold value of the proximity factor. The dumb node $D$ estimates its position $(x_{D1}, y_{D1})$ at the center of these selected beacon nodes as under:

$$(3) \quad x_{D1} = \frac{x_{B1} + ... + x_{BN1}}{N_1}$$

$$(4) \quad y_{D1} = \frac{y_{B1} + ... + y_{BN}}{N_1}$$

The dumb node $D$, then selects another set of 3 or more beacon nodes whose proximity factor falls in the next highest range of threshold values. The dumb node $D$ then makes another estimate of its location $(x_{D2}, y_{D2})$ at the center of this new set of beacon nodes. The dumb node repeats this step until all sets of beacon nodes in its table of beacon nodes are exhausted. Finally, the dumb node $D$ estimates its position as the weighted average of the positions calculated above:

$$(5) \quad x_D = \frac{F_{M1} x_{D1} + ... + F_{MN} x_{DN}}{F_{M1} + ... + F_{MN}}$$

$$(6) \quad y_D = \frac{F_{M1} y_{D1} + ... + F_{MN} y_{DN}}{F_{M1} + ... + F_{MN}}$$

$F_{M1}$ is the mean proximity factor for the first threshold value of proximity factor, $F_{M2}$ is the mean proximity factor for the second threshold and so on.

Only a dumb node with 3 or more beacon nodes in its neighborhood determines its position using the above steps. After the dumb node localizes itself, it becomes a beacon node and starts transmitting beacon signals. As a result, another beacon node becomes available in the sensor field. From the remaining dumb nodes, again dumb nodes with 3 or more beacon nodes in their neighborhood localize themselves using above steps and then become available as beacon nodes. The above steps are repeated until all dumb nodes with three or more neighbor beacon nodes localize themselves.

The algorithm is described below using pseudo code.

```
// Beacon Nodes
while ( count < totalBeaconNodes )
  reset (timer);
  start (timer);
  select beaconNode(count);
```

```
            prepare ( beaconMessage);
            add beaconPosition=>beaconMessage;
            if (timer == T )
                     transmit (beaconMessage);
            end if
      end select
end while

// Dumb Nodes
receive (beaconMessage);
get beaconPosition => beaconMessage;
if (! lookup(beaconPosition), beaconTable )
     add (beaconPosition, beaconTable);
     beaconCount++;
end if

while( beacon < beaconCount )
     reset timer;
     start timer;
     while ( timer < sampleTime )
            if (receive (beacon) == TRUE )
                     receivedBeacons++;
            end if
     end while
     totalBeacons = sampleTime/T;

     proximity factor = F_k = receivedBeacons / totalBeacons;
     add proximity factor => beaconTable;
end while

prepare (list of threshold values);
N = number of threshold values;
for ( k=0; k<N, k++ )
     select (nodes n >= 3 from beaconTable)
            Xk = ( X_{b1} + ... + X_{bn} )/ n;
            Yk = ( Y_{b1} + ... + Y_{bn} )/ n;
     end select
end for

X_d = ( F_1X_1 + ... + F_NX_N ) / (F_1 + ... + F_N);
Y_d = ( F_1Y_1 + ... + F_NY_N ) / (F_1 + ... + F_N);
convert ( dumbNode => beaconNode);

end
```

The algorithm uses only connectivity information to infer the position estimates of the dumb nodes and hence is range free. For the most part, the algorithm is passive i.e. a dumb node localizes itself using only the received messages. Furthermore, no additional hardware is required to be added or installed on the nodes for localization. As a result, minimum energy is consumed and the algorithm is energy-efficient.

**Simulation and results**

To evaluate the localization algorithm, it was simulated using Omnet++. A number of simulation experiments were carried out. The results were analyzed using the Result Analysis Tool provided with Omnet++ and were plotted using Gnuplot.

*Simulation setup*

For simulation of the algorithm for a wireless sensor network, composition of the sensor node and topology of sensor network is described using network description (NED) language. Further, the functioning of a node and the localization algorithm running on it is described using C/C++.

Simulation experiments are performed on randomly deployed sensor nodes in a rectangular sensor field. Size of the sensor field is 100 m x 100 m, and a total of 400 sensor nodes are placed randomly in the sensor field. Each sensor node has a radio range of 10 m. The simulation parameters are summarized in Table 1.

Table 1. Simulation parameters

| Simulation Parameter | Value |
| --- | --- |
| Number of nodes | 400 |
| Length of sensor field | 100 m |
| Width of sensor field | 100 m |
| Radio range of a node | 10 m |

Important results from the simulation experiments are summarized in the following sections.

*Localization error*

When the sensor nodes are randomly deployed in the sensor field, their actual positions are recorded. After the localization algorithm is run, their estimated positions are also recorded. The distance between the actual and estimated positions gives error in localization or simply it is termed as localization error. If the actual position of a sensor node is $(X, Y)$ and the estimated position is $(X_e, Y_e)$, then the error in localization i.e. the distance between the actual and estimated positions is given by:

$$(7) \qquad e = \sqrt{(X - X_e)^2 + (Y - Y_e)^2}$$

The error $e$ is then normalized by dividing it with the radio range $R$ of the sensor node.

Fig. 1 shows a plot of localization error for the first 200 nodes in the sensor field. It is evident from the graph that the mean localization error for the simulation experiment is below 0.5. Considering the fact that the localization algorithm is range free, it produced better than coarse-grained location estimate on average. The error mode is almost 0.3 which indicates the fact that the localization error of majority of nodes is below the mean localization error. There are only a few peaks above localization error of 1, which results in a mean error greater than the error mode. The standard deviation in error is also low indicating that the variation in localization errors of different nodes is small. If the localization error result is interpreted by the combined values of error mode and error standard deviation, it means that majority of sensor nodes having modal error of approximately 0.3 and have only small variations in this modal error.

*Error distribution*

The graph in Fig. 2 is a CDF of the localization error. It tells us the distribution of error in the sensor nodes i.e. number of nodes having a certain value of localization error. The graph confirms the result in Fig. 1. More than 90% of the sensor nodes have localization error below the mean error of 0.5 and only 20% of sensor nodes have localization error greater than 0.2.

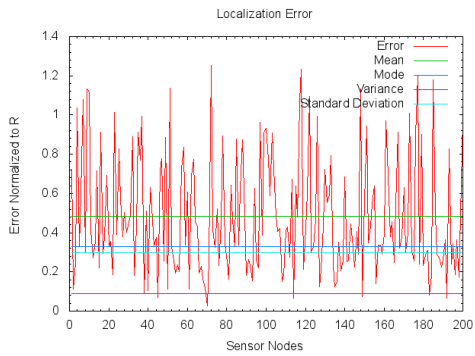

Fig. 1. Error in localization.

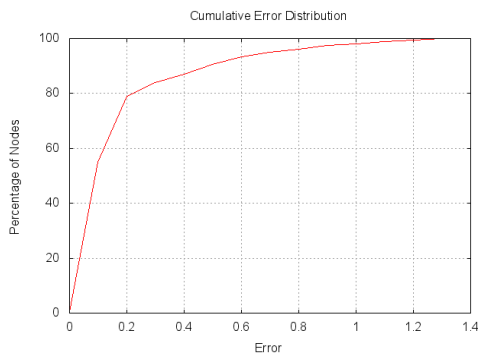

Fig. 2. Error distribution.

*Geographic distribution of error*

The graph in Fig. 3 shows the amount of localization error at various positions in the sensor field. It is evident from the graph that the localization error is high at the boundaries and corners of the sensor field. This is due to the fact that the localization algorithm tends to localize a sensor node at the *center* of selected beacon nodes. The dumb nodes at the boundaries and corners are not surrounded by other nodes on all sides. Therefore, when a node tries to localize itself using other nodes which do not surround the dumb node and are rather positioned only on its one or two sides, the algorithm tends to localize the node away from its original position towards the helping nodes.

*Error statistics and number of nodes*

A number of simulation experiments with different number of total nodes were carried out and results for the localization error were recorded. The results were then analyzed and the mean, mode, variance and standard deviation of error were calculated. Summary of the results are shown in Table II and are plotted in Fig. 4. The results for the error mode are interesting. The error mode was calculated as a single decimal value because it gives better accuracy of number of nodes having a particular localization error. It is to be noted that the results for error mode are only an approximation. It is for this reason that error mode for 150 nodes is 0.1 and for 200 nodes it is 0.2 whereas one would expect it to be lower than 0.1 when the number of nodes is increased from 150 to 200. The increase in error mode may be due to the fact that with the increase in total number of nodes from 150 t0 200, now more nodes are available whose results are improved from localization error of, say 0.3, to 0.2. Furthermore, the results should be seen combined with the mean localization error which has reduced. Lower values of variance and standard deviation also depict improvement in results.

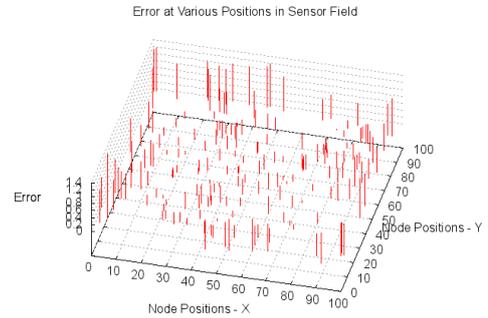

Fig. 3. Geographic distribution of error

Table 2. Error Statistics and Number of Nodes

| Total Nodes | Mean Error | Mode | Variance | Standard Deviation |
|---|---|---|---|---|
| 100 | 0.81 | 0.6 | 0.1679 | 0.4098 |
| 150 | 0.74 | 0.1 | 0.1477 | 0.3843 |
| 200 | 0.66 | 0.2 | 0.1446 | 0.3803 |
| 250 | 0.64 | 0.1 | 0.1183 | 0.3439 |
| 300 | 0.60 | 0.2 | 0.1075 | 0.3279 |
| 350 | 0.53 | 0.1 | 0.0976 | 0.3124 |
| 400 | 0.53 | 0.1 | 0.1040 | 0.3226 |

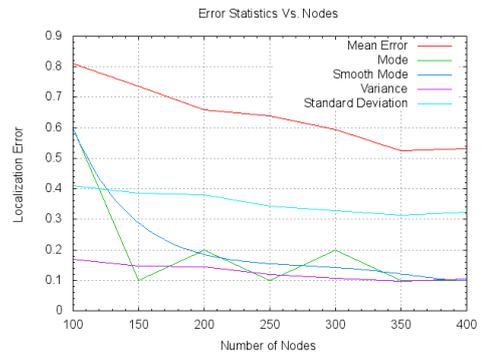

Fig. 4. Error statistics and number of nodes

*Blind nodes*

Some of the sensor nodes in the sensor field may not be surrounded by sufficient number of other nodes at hearing distance from them. For example, a node at a boundary of the sensor field may have only two neighbor nodes at hearing distance from it. Under these conditions, the node will not be able to localize itself and is termed as a blind node.

Graph in Figure 5 shows the number of blind nodes as the total number of nodes in the sensor field is increased. As the number of nodes increases, more neighbor nodes are available to a dumb node for localization. Therefore, the number of blind nodes reduces with the increase of number of nodes in the sensor field.

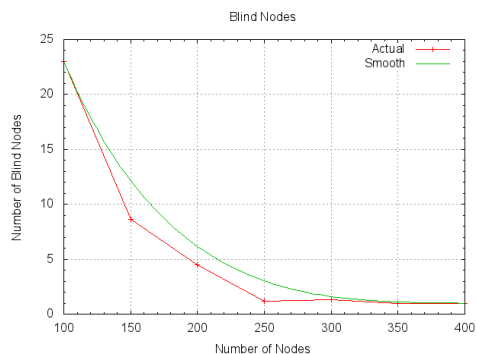

Fig. 5. Blind nodes

*Summary of results*

From the results described above, it can be seen that majority of sensor nodes are able to localize themselves effectively. However, a closer observation reveals that some of the sensor nodes at the boundaries and corners of the square shaped sensor field are not able to localize. The nodes at these positions may remain blind. The problem can be alleviated if the area of investigation is managed to be in the center of the sensor field.

**Conclusion**

In this paper, we have defined and formulated the problem of localization of nodes in wireless sensor networks and have also discussed a few of the representative localization algorithms that are closer in nature to our proposed algorithm. We have also presented a connectivity based localization algorithm which is well suited to the distributed ad hoc nature of wireless sensor networks. The simulation results show that the localization algorithm is able to achieve good location accuracy.

REFERENCES

[1] De Oliveira, H. A. B. F., Nakamura, F. E., Loureiro, A. A. & Boukerche, A. (2005, October). Directed Position Estimation: A Recursive Localization Approach for Wireless Sensor Networks. In *Proceedings of the 14th International Conference on Computer Communications and Networks*, ICCCN '05, pp. 557–562.

[2] Want, R., Falcao, V., & Gibbons, J. (1992, January). The Active Badge Location System, *ACM Transactions on Information Systems*, 10(3).

[3] Harter, A., & Hopper, A. (1994, January). A Distributed Location System for the Active Office, *IEEE Network*, 8(1), 62-70.

[4] Ward, A., Jones, A., & Hopper, A. (1997, October). A New Location Technique for the Active Office, *IEEE Personal Communications*, 4(5), 42 – 47.

[5] Priyantha, N. B. (2005). The Cricket Indoor Location System. *Doctoral Dissertation*, Massachusetts Institute of Technology, Cambridge, USA.

[6] Priyantha, N. B., Chakraborty, A., & Balakrishnan, H. (2000). The Cricket Location-Support System, In *Proceedings of the 6th Annual International Conference on Mobile Computing and Networking*, Mobicom '00, ACM.

[7] Priyantha, N. B., Miu, A. K., Balakrishnan, H., & Teller, S. (2001, July). The Cricket Compass for Context-Aware Mobile Applications. In *Proceedings of the 7th Annual International Conference on Mobile Computing and Networking*, Mobicom '01.

[8] Bulusu, N., Heidemann, J., & Estrin, D. (2000, October). GPS-less Low-Cost Outdoor Localization for Very Small Devices. *IEEE Personal Communications*, pp. 28-34, IEEE Communications Society.

[9] He, T., Huang, C., Blum, B., Stankovic, J. A., & Abdelzaher, T. (2003). Range-Free Localization Schemes for Large Scale Sensor Networks, In *Proceedings of the 9th Annual International Conference on Mobile Computing and Networking*, Mobicom '03, (pp. 535-539).

[10] Albowicz, J., Chen, A., & Zhang, L. (2001). Recursive PositionEstimation in Sensor Networks. In *Proceedings of the 9th International Conference on Network Protocols*, ICNP 2001, pp. 35–41.

[11] De Oliveira, H. A. B. F., Nakamura, E. F., Loureiro, A. A. F., & Boukerche, A. (2007, May). Localization in Time and Space for Sensor Networks. In *Proceedings of 21st International Conference on Advanced Information Networking and Applications Workshops*, AINAW'07, pp. 539-546, IEEE Computer Society.

[12] Doherty, L., Pister, K. S. J., & Ghaoui, L. E. (2001, April). Convex Position Estimation in Wireless Sensor Networks. In *Proceedings of Twentieth Annual Joint Conference of the IEEE Computer and Communications Societies*, INFOCOM 2001, Vol. 3. (pp. 1655-1663). IEEE Computer Society.

[13] Kim, E., & Kim., K. (2010, June). Distance Estimation With Weighted Least Squares for Mobile Beacon-Based Localization in Wireless Sensor Networks. *IEEE Signal Processing Letters*, Vol. 7, Issue 6, pp. 559-562, IEEE Signal Processing Society.

[14] Niculescu, D., & Nath, B. (2001). Ad Hoc Positioning System, In *Proceedings of the IEEE Global Telecommunications Conference*, GLOBECOM '01, pp. 2926-2931, IEEE.

[15] Savarese, C., Rabaey, J. M., & Beutel, J. (2001, May). Locationing in Distributed Ad-Hoc Wireless Sensor Networks. In *Proceedings of the International Conference on Acoustics, Speech and Signal Processing*, ICASSP 2001, Salt Lake City, UT, USA, (pp. 2037-2040).

[16] Savarese, C., Rabaey, J. M. & Langendoen, K. (2002). Robust Positioning Algorithms for Distributed Ad-Hoc Wireless Sensor Networks. In *Proceedings of the General Track: 2002 USENIX Annual Technical Conference*. pp. 317–327, Berkeley, CA, USA: USENIX Association.

[17] Savvides, A., Han, C. C., & Srivastava, M. B. (2001, July). Dynamic Fine-Grained Localization in Ad-Hoc Networks of Sensors. In *Proceedings of the 7th Annual International Conference on Mobile Computing and Networking*, Mobicom'01, (pp. 166-179).

[18] Savvides, A., Park, H., & Srivastava, M. B. (2002, September). The Bits and Flops of the N-hop Multilateration Primitive For Node Localization Problems. In *Proceedings of the 1st ACM International Workshop on Wireless Sensor Networks and Applications*, WSNA '02, pp. 112-121, ACM Press.

[19] Shang, Y., & Ruml, W. (2004, March). Improved MDS-Based Localization. In *Proceedings of Twenty Third Annual Joint Conference of the IEEE Computer and Communications Societies*, IEEE INFOCOM 2004 (vol. 4 pp. 2640-2651).

[20] Shang, Y., Ruml, W., Zhang, Y., & Fromherz, M. P. J. (2003, June). Localization from Mere Connectivity. In *Proceedings of the 4th ACM International Symposium on Mobile Ad Hoc Networking & Computing*, MobiHoc'03 (pp. 201-212). ACM Press.

[21] Shang, Y., Ruml, W., Zhang, Y., & Fromherz, M. (2004, October). Localization from Connectivity in Sensor Networks. *IEEE Transactions on Parallel and Distributed Systems*, 15(11), 961-974.

[22] Chang, C. Y., Hung, L. L., Lin, C. Y., & Li, M. H. (2010, May). On Distinguishing Relative Locations with Busy Tones for Wireless Sensor Networks. In *Proceedings of The 2010 IEEE International Conference on Communication*, (pp. 1-5).

[23] Wang, C. L., Wu D. S., & Shu, F. F. (2010, April). Design and Implementation of Decentralized Positioning System for Wireless Sensor Networks. In *Proceedings of The 2010 IEEE Wireless Communications and Networking Conference*, (pp. 1-6).